\documentstyle[camera,prbplug,psfig]{article}

\begin{document}

\title{ {\small JOURNAL OF THE PHYSICAL SOCIETY OF JAPAN 67 (1998) 3867-3874}\\
\vspace{0.7cm}
Off-diagonal Wave Function Monte Carlo Studies of Hubbard Model I}
\author{Takashi Yanagisawa, Soh Koike and Kunihiko Yamaji}
\address{$^a$Electrotechnical Laboratory, 1-1-4 Umezono, Tsukuba, Ibaraki 
305-8568, Japan}

\maketitle
 
\begin{abstract}
We propose a Monte Carlo method, which is a hybrid method of the quantum
Monte Carlo method and variational Monte Carlo theory,
to study the Hubbard model.
The theory is based on the off-diagonal and the Gutzwiller type correlation 
factors which are taken into account by a Monte Carlo algorithm.
In the 4$\times$4 system our method is able to reproduce the exact 
results obtained by the diagonalization.  
An application is given to investigate the half-filled band case of 
two-dimensional square lattice.
The energy is favorably compared with
quantum Monte Carlo data.
\end{abstract}

\section{Introduction}
Strongly correlated electron systems have been investigated by many
researchers in order to understand the mechanism of superconductivity of
the high-T$_{\rm c}$ oxide superconductors.  The effect of strong correlation
between electrons is important for high-T$_{\rm c}$ superconductivity and
many unconventional phenomena such as metal-insulator transition and heavy
fermions with a huge large mass.  The Hubbard model is a
basic model for strongly correlated electrons in metals.
The Hubbard model has been regarded as a model to describe the Mott 
transition in compounds such as NiO and MnO.\cite{mot74}
It is considered that the Hubbard model contains basic physics of high-T$_c$
superconductivity as a simplified one-band model of the three band Cu-O
model.\cite{and87} 
A possibility of superconductivity for the 2D Hubbard model has been
controversial for many years.\cite{hir85,whi89,sca90}
The study of the Hubbard model provides us insights having important 
implications concerning the origin of high-T${\rm c}$ superconductivity.
The possibility of the superconducting state is recently reported for the
Hubbard model\cite{gia91,nak97,kur97,yam98} and the two-chain Hubbard 
model.\cite{yam94,yan95,kur96}
The phase diagram is still far from well understood in two dimensions.

The quantum Monte Carlo method is a method to treat the strong correlations
exactly.  However, the applicability is restricted to moderately
correlated region because of a sign problem.
The variational Monte Carlo method has a feature characterized by a wide
applicability from weak to strong correlation.
The Gutzwiller wave function is a standard trial wave function for itinerant
correlated electrons.  In the Gutzwiller function only the correlation at
the same site is taken into account and thus the wave function is very
simple in its form.
An analytic evaluation of expectation values is, however, very difficult
for the Gutzwiller function in spite of its simplicity.
This difficulty is overcome by a Monte Carlo method
and many calculations have been performed using the Monte Carlo
algorithm.\cite{gia91,cep77,yok87,gro87}

It is revealed that the Gutzwiller function will give higher energies
compared to exact values. Thus we think that the Gutzwiller 
function should be improved
to investigate the ground state more precisely.
The purpose of this paper is to propose a hybrid method of the quantum Monte
Carlo (QMC) calculations and variational Monte Carlo (VMC) theory.
Following an idea of QMC, a trial function is improved and the expectation
values of energy and other quantities are calculated using VMC for the
parameters optimizing the energy expectation value. 
The off-diagonal wave function Monte Carlo method (OWMC) in this paper
deal with
Gutzwiller projection and off-diagonal correlation operators taken into
account by a Monte Carlo algorithm.  We expect to be able to extrapolate 
correct expectation values
from the data obtained for off-diagonal wave functions.

We show advantages of OWMC in the following: (1) The calculations give an
upper bound of the exact energy since OWMC is based on the variational theory.
(2) Variational wave functions are improved systematically.
(3) There is no sign problem.  (4)  The large-$U$ systems are tractable.
(5) Introducing the order parameters we can investigate ground state
properties characterized by the long-range ordering.

The paper is organized as follows.  The section 2 is devoted to present the
Hamiltonian and wave functions.  The method of calculations is also shown.
The following two sections are assigned to a description of results.  The 
summary and discussion are presented in the final section.

\section{Hamiltonian and wave functions}
 
\subsection{Wave functions}

The Hamiltonian is given by the Hubbard model:

\begin{equation}
H = -t\sum_{\langle ij\rangle}(c^{\dag}_{i\sigma}c_{j\sigma}+{\rm h.c.})
+U\sum_i n_{i\uparrow}n_{i\downarrow},
\end{equation}
where $n_{i\sigma}=c^{\dag}_{i\sigma}c_{i\sigma}$ is the number operator and
$U$ is the strength of on-site Coulomb interaction.
The energy is measured in units of $t$ throughout this paper and the number
of sites is denoted as $N$.
In this paper the Hubbard model is considered in two space dimensions.
The Coulomb interaction $U$ is expected to bring about
the metal-insulator transition, antiferromagnetic order and a
possibility of the anisotropic superconducting state.

Let us start with the Gutzwiller function given by
\begin{equation}
\psi_G = P_G\psi_0 =
{\rm exp}(-\alpha\sum_in_{i\uparrow}n_{i\downarrow})\psi_0
\equiv \psi^{(0)},
\end{equation}
where $\psi_0$ is the free fermion ground state.
$P_G$ is the Gutzwiller operator given by
$P_G=\prod_i (1-(1-g)n_{i\uparrow}n_{i\downarrow})$ where $g$ is the
variational parameter in the range of $0\leq g\leq 1$ and 
$\alpha={\rm log}(1/g)$.
Our method is based on the fact that
the ground state eigenfunction is written as in QMC
\begin{equation}
\psi = e^{-\beta H}\psi_0  \simeq 
e^{-\epsilon_1 K}e^{-\epsilon_1 V'}\cdots e^{-\epsilon_n K}e^{-\epsilon_n V'}
\psi_0,  \label{wavef}
\end{equation}
for large $\beta=\epsilon_1+\cdots+\epsilon_n$ and small $\epsilon_i$
($i=1,\cdots,n$).  
$H$ is written as $H=K+V'$
where $K$ denotes the kinetic energy part and $V'=UV$ denotes the
on-site interaction part: $V=\sum_in_{i\uparrow}n_{i\downarrow}$.
Since the last factor $e^{-\epsilon V'}\psi_0$ in eq.(\ref{wavef}) is
regarded as the Gutzwiller function, one way to improve $\psi_G$ is
achieved by multiplying $e^{-\lambda K}$ to $\psi_G$.  Therefore we can
consider\cite{oht92}
\begin{equation}
\psi^{(1)} = e^{-\lambda K}e^{-\alpha V}\psi_0.
\end{equation}
It has been shown that the Gutzwiller function is improved appreciably
by the off-diagonal correlation factor $e^{-\lambda K}$.\cite{oht92,yan98,yan98b}
The next step is to multiply $e^{-\alpha' V}$ in order to control the double
occupancy in $\psi^{(1)}$.  We can again operate $e^{-\lambda K}$
to improve the wave function.  A second-level wave function is given by
\begin{equation}
\psi^{(2)} = e^{-\lambda' K}e^{-\alpha' V}e^{-\lambda K}
e^{-\alpha V}\psi_0,
\end{equation}
where $\lambda$, $\lambda'$, $\alpha$ and $\alpha'(={\rm log}(1/g'))$ are
variational parameters.  Next wave function is written as
\begin{equation}
\psi^{(3)} = e^{-\lambda'' K}e^{-\alpha'' V}e^{-\lambda' K}
e^{-\alpha' V}e^{-\lambda K}e^{-\alpha V}\psi_0.
\end{equation}  
$\lambda$, $\lambda'$, $\lambda''$, $\alpha$, $\alpha'$ and $\alpha''$ are
variational parameters. 
It is considered that $\psi^{(m)}$ with optimized variational parameters 
approaches the ground state wave function
in eq.(3) as $m$ grows larger.  Although a sign problem will occur for large
$m$, the sign never brings about a problem for small $m$.
If we can extrapolate the expectation values from the data obtained using
$\psi^{(1)}$, $\psi^{(2)}$, $\cdots $,  we can estimate exact values within
Monte Carlo errors.

\subsection{Method of calculations}

A Monte Carlo algorithm developed in the auxiliary-field quantum Monte
Carlo calculations\cite{bla81} enables us to evaluate the expectation
values for the wave functions $\psi^{(1)}$, $\psi^{(2)}$, $\cdots$.
Using the discrete Hubbard-Stratonovich transformation, the Gutzwiller factor is
written as
\onecolumn
\begin{equation}
{\rm exp}(-\alpha\sum_i n_{i\uparrow}n_{i\downarrow})=
(1/2)^N \sum_{s_i}{\rm exp}[2a\sum_is_i(n_{i\uparrow}
-n_{i\downarrow})
- \frac{\alpha}{2}\sum_i(n_{i\uparrow}+n_{i\downarrow})],
\end{equation}
where $a$ is given by ${\rm cosh}(2a)=e^{\alpha/2}$.
$N$ denotes the number of sites.
The auxiliary fields $s_i$ takes the values of $\pm 1$.
The norm $\langle\psi_G|\psi_G\rangle$ is written as
\begin{equation}
\langle\psi_G|\psi_G\rangle = (1/2)^{2N}
\sum_{\{u_i\}\{s_i\}}\prod_{\sigma}\langle\psi^{\sigma}_0|
{\rm exp}(h^{\sigma}(u)){\rm exp}(h^{\sigma}(s))|\psi^{\sigma}_0\rangle ,
\end{equation}
where the potential $h^{\sigma}(u)$ is given by
\begin{equation}
h^{\sigma}(u)= 2a\sigma\sum_iu_in_{i\sigma}-\frac{\alpha}{2}\sum_in_{i\sigma}.
\end{equation}
Then the weight is written as a sum of determinants,\cite{ima89,zha97}

\begin{equation}
\langle\psi_G|\psi_G\rangle ={\rm const.} (1/2)^{2N}\sum_{\{u_i\}\{s_i\}}\prod_{\sigma}
{\rm det}(\phi^{\sigma\dag}_0{\exp}(V^{\sigma}(u,\alpha)){\rm exp}(V^{\sigma}(s,\alpha))
\phi^{\sigma}_0),
\end{equation}
where $V^{\sigma}(s,\alpha)$ is a diagonal $N\times N$ matrix corresponding
to $h^{\sigma}(s)$ written as
$V^{\sigma}(s,\alpha)={\rm diag}(2a\sigma s_1-\alpha/2,\cdots,2a\sigma s_N-\alpha/2)$, where diag($a,\cdots$) denotes a diagonal matrix with elements 
given as the arguments $a,\cdots$.
For the free fermion state, the elements of $\phi^{\sigma}_0$ are given by 
plane waves:
$(\phi^{\sigma}_0)_{ij}={\rm exp}(i{\bf r}_i\cdot {\bf k}_j)$
($i=1,\cdots,N; j=1,\cdots,N_e/2$ where $N_e$ is the number of electrons).  
We may take ${\bf k}_j$ below the Fermi surface corresponding to the free
fermion ground state as a trial state.  To represent $\phi^{\sigma}_0$ as a
real matrix, we take $(\phi^{\sigma}_0)_{ij}$ as 
${\rm cos}({\bf r}_i\cdot {\bf k}_j)$ or ${\rm sin}({\bf r}_i\cdot {\bf k}_j)$.
A more general choice is possible by incorporating the spin-density wave (SDW)
long range order.
The one-particle antiferromagnetic ordered state is given by
\begin{equation}
\psi_{{\rm SDW}} = \prod_{\bf k}(u_{\bf k}c^{\dag}_{{\bf k}\uparrow}+v_{\bf k}c^{\dag}_{{\bf k}+{\bf Q}\uparrow})
\prod_{{\bf k}'}(u_{{\bf k}'}c^{\dag}_{{\bf k}'\downarrow}-v_{{\bf k}'}c^{\dag}_{{\bf k}'+{\bf Q}\downarrow})
|0\rangle ,
\end{equation}
where $u_{\bf k}=[(1-\epsilon_{\bf k}/(\epsilon_{\bf k}^2+\Delta_{AF}^2)^{1/2})/2]^{1/2}$ and
$v_{\bf k}=[(1+\epsilon_{\bf k}/(\epsilon_{\bf k}^2+\Delta_{AF}^2)^{1/2})/2]^{1/2}$.
$\Delta_{AF}$ is the antiferromagnetic order parameter optimizing the energy.
${\bf Q}$ denotes SDW wave vector given as ${\bf Q}=(\pi,\pi)$.
The elements of $\phi^{\sigma}_0$ corresponding to $\psi_{{\rm SDW}}$ are 
written as
\begin{equation}
(\phi^{\sigma}_0)_{ij}= u_{{\bf k}_j}{\rm exp}(i{\bf r}_i\cdot {\bf k}_j)
+ {\rm sign}{\sigma}v_{{\bf k}_j}{\rm exp}(i{\bf r}_i\cdot ({\bf k}_j+{\bf Q})).
\end{equation}
In real representation they are given by
$u_{{\bf k}_j}{\rm cos}({\bf r}_i\cdot {\bf k}_j)$
+${\rm sign}{\sigma}v_{{\bf k}_j}{\rm cos}({\bf r}_i\cdot ({\bf k}_j+{\bf Q}) )$
and
$u_{{\bf k}_j}{\rm sin}({\bf r}_i\cdot {\bf k}_j)$
+${\rm sign}{\sigma}v_{{\bf k}_j}{\rm sin}({\bf r}_i\cdot ({\bf k}_j+{\bf Q}) )$. 

Similarly the norms including
the off-diagonal factors are written as

\begin{equation}
\langle\psi^{(1)}|\psi^{(1)}\rangle ={\rm const.}(1/2)^{2N}\sum_{\{u_j\}\{s_i\}}
\prod_{\sigma}
{\rm det}(\phi^{\sigma\dag}_0{\rm exp}(V^{\sigma}(u,\alpha))
{\rm exp}(-\lambda K^{\sigma}){\rm exp}(-\lambda K^{\sigma})
 {\rm exp}(V^{\sigma}(s,\alpha))\phi^{\sigma}_0),
\end{equation}
and
\begin{eqnarray}
\langle\psi^{(2)}|\psi^{(2)}\rangle &=& {\rm const.}(1/2)^{4N}\sum_{\{u_j\}\{v_{\ell}\}\{t_k\}\{s_i\}}
\prod_{\sigma}
{\rm det}(\phi^{\sigma\dag}_0{\rm exp}(V^{\sigma}(u,\alpha))
{\rm exp}(-\lambda K^{\sigma})
{\rm exp}(V^{\sigma}(v,\alpha'))\nonumber\\
&\times& {\rm exp}(-\lambda' K^{\sigma}) {\rm exp}(-\lambda' K^{\sigma})
{\rm exp}(V^{\sigma}(t,\alpha'))
{\rm exp}(-\lambda K^{\sigma})
{\rm exp}(V^{\sigma}(s,\alpha))\phi^{\sigma}_0),
\end{eqnarray}
\twocolumn
where $K^{\sigma}$ is a matrix corresponding to the kinetic part of the
Hamiltonian:

\begin{eqnarray}
(K^{\sigma})_{ij}&=& -t~~{\rm if}~(i,j)~{\rm are}~ {\rm nearest}~ {\rm neighbor}~{\rm sites},\nonumber\\
&=& 0~~~{\rm otherise}.
\end{eqnarray}
 
For two states $|\psi^{\sigma}\rangle$ and $|\tilde{\psi}^{\sigma}\rangle$ represented
by Slater determinants $\phi^{\sigma}$ and $\tilde{\phi}^{\sigma}$ as defined 
above respectively,
the single-particle Green function is written as\cite{zha97}
\begin{eqnarray}
\langle\psi^{\sigma}|c_{i\sigma}c^{\dag}_{j\sigma}|\tilde{\psi}^{\sigma}\rangle&/&
\langle\psi^{\sigma}|\tilde{\psi}^{\sigma}\rangle = \delta_{ij}\nonumber\\
&-&
(\tilde{\phi}^{\sigma}(\phi^{\sigma\dag}\tilde{\phi}^{\sigma})^{-1}\phi^{\sigma\dag})_{ij},
\end{eqnarray}
for $\langle\psi^{\sigma}|\tilde{\psi}^{\sigma}\rangle\neq 0$.

In order to evaluate the expectation value we generate the Monte Carlo
samples by the importance sampling with the weight function 
$|w|$=$|w_{\uparrow}w_{\downarrow}|$ where
\begin{equation}
w_{\sigma}= {\rm det}(\phi^{\sigma\dag}_0{\rm exp}(V^{\sigma}(u,\alpha))\cdots
{\rm exp}(V^{\sigma}(s,\alpha))\phi^{\sigma}_0).
\end{equation}
When we update the Ising variable from the old $s_i$ to the new one $s'_i$,
the ratio of $|w_{\uparrow}w_{\downarrow}|$ is calculated to determine
whether to accept or reject the new configuration.
In the process of updating $s_i$ to $s'_i$,
the Gutzwiller factor ${\rm exp}(V^{\sigma}(s,\alpha))$ is written as
\onecolumn
\begin{eqnarray}
{\rm exp}(V^{\sigma}(s_1,\cdots,s'_i,\cdots,s_N,\alpha))&=&
{\rm diag}({\rm e}^{2a\sigma s_1-\alpha/2},\cdots,{\rm e}^{2a\sigma s'_i-\alpha/2},\cdots,
{\rm e}^{2a\sigma s_N-\alpha/2})\nonumber\\
&=& (1+\Delta_{\sigma}){\rm exp}(V^{\sigma}(s_1,\cdots,s_i,\cdots,s_N,\alpha)),
\end{eqnarray}
where $\Delta_{\sigma}$ is a diagonal matrix given by
$\Delta_{\sigma}={\rm diag}(0\cdots 0,{\rm e}^{2a\sigma(s'_i-s_i)}-1,0\cdots )$.
Then the ratio of $|w_{\sigma}|$ is given by\cite{ima89}
\begin{eqnarray}
r_{\sigma}&=& |{\rm det}(\phi^{\sigma\dag}_0{\rm exp}(V^{\sigma}(u,\alpha))
\cdots (1+\Delta_{\sigma}){\rm exp}(V^{\sigma}(s,\alpha))\phi^{\sigma}_0)|/
|{\rm det}(\phi^{\sigma\dag}_0{\rm exp}(V^{\sigma}(u,\alpha))\cdots
{\rm exp}(V^{\sigma}(s,\alpha))\phi^{\sigma}_0)|\nonumber\\
&\equiv& |{\rm det}(L(1+\Delta_{\sigma})R|/|{\rm det}(LR)|\nonumber\\
&=& |{\rm det}(1+L\Delta_{\sigma}RJ)|
= |1+(\Delta_{\sigma})_{ii}(G_{r\sigma})_{ii}|,
\end{eqnarray}
where the right Green function is written as
\begin{equation}
G_{r\sigma}= {\rm exp}(V^{\sigma}(s,\alpha))\phi^{\sigma}_0 J\phi^{\sigma\dag}_0{\rm exp}(V^{\sigma}(u,\alpha)){\rm exp}(-\lambda K^{\sigma})\cdots
{\rm exp}(-\lambda K^{\sigma}),
\end{equation}
with 
\begin{equation}
J = \{\phi^{\sigma\dag}_0{\rm exp}(V^{\sigma}(u,\alpha))
{\rm exp}(-\lambda K^{\sigma})\cdots {\rm exp}(-\lambda K^{\sigma})
{\rm exp}(V^{\sigma}(s,\alpha))\phi^{\sigma}_0\}^{-1}.
\end{equation}
\twocolumn
The updated Green function is straightforwardly calculated from the old one
if the new variable $s'_i$ is accepted:
\begin{equation}
G^{new}_{r\sigma}= (1+\Delta_{\sigma})[G_{r\sigma}-G_{r\sigma}\Delta_{\sigma}
(1+G_{r\sigma}\Delta_{\sigma})^{-1}G_{r\sigma}].
\label{newgreen}
\end{equation}
We continue the updating procedure for other Ising variables.
The relation in eq.(\ref{newgreen}) is derived as follows.
$G^{new}_{r\sigma}$ is written as
\onecolumn
\begin{equation}
G^{new}_{r\sigma}= (1+\Delta_{\sigma}){\rm exp}(V^{\sigma}(s,\alpha))
\phi^{\sigma}_0 J'\phi^{\sigma\dag}_0{\rm exp}(V^{\sigma}(u,\alpha))\cdots 
{\rm exp}(-\lambda K^{\sigma}) \equiv 
(1+\Delta_{\sigma})\tilde{G}^{new}_{r\sigma},
\label{newdef}
\end{equation}
where $J'$ is given by
\begin{equation}
J'= \{\phi^{\sigma\dag}_0{\rm exp}(V^{\sigma}(u,\alpha))
{\rm exp}(-\lambda K^{\sigma})\cdots {\rm exp}(-\lambda K^{\sigma})
{\rm exp}(V^{\sigma}(s_1,\cdots,s'_i,\cdots,s_N,\alpha))\phi^{\sigma}_0\}^{-1}.
\end{equation}
Inserting the relation
\begin{eqnarray}
1&=& J\phi^{\sigma\dag}_0{\rm exp}(V^{\sigma}(u,\alpha))
{\rm exp}(-\lambda K^{\sigma})\cdots {\rm exp}(-\lambda K^{\sigma})
{\rm exp}(V^{\sigma}(s,\alpha))\phi^{\sigma}_0\nonumber\\
&=& J[ (J')^{-1}-\phi^{\sigma\dag}_0{\rm exp}(V^{\sigma}(u,\alpha))
{\rm exp}(-\lambda K^{\sigma})\cdots {\rm exp}(-\lambda K^{\sigma})
\Delta_{\sigma}{\rm exp}(V^{\sigma}(s,\alpha))\phi^{\sigma}_0 ],
\end{eqnarray}
into the left of J' in eq.(\ref{newdef}),
$\tilde{G}^{new}_{r\sigma}$ is given as
\begin{equation}
\tilde{G}^{new}_{r\sigma}= G_{r\sigma}-G_{r\sigma}\Delta_{\sigma}
\tilde{G}^{new}_{r\sigma}.
\end{equation}
Then the eq.(\ref{newgreen}) is followed.
When we update the variable $u_i$ in the left part of $w_{\sigma}$ to new one 
$u'_i$, the updated Green function is similarly calculated as
\begin{equation}
G^{new}_{\ell\sigma}= [G_{\ell\sigma}-
G_{\ell\sigma}(1+\Delta_{\sigma}G_{\ell\sigma})^{-1}\Delta_{\sigma}G_{\ell\sigma}](1+\Delta_{\sigma}),
\end{equation}
where the old Green function is given as
\begin{equation}
G_{\ell\sigma}= {\rm exp}(-\lambda K^{\sigma})\cdots 
{\rm exp}(-\lambda K^{\sigma}){\rm exp}(V^{\sigma}(s,\alpha))\phi^{\sigma}_0
J\phi^{\sigma\dag}_0{\rm exp}(V^{\sigma}(u,\alpha)),
\end{equation}
and $\Delta_{\sigma}$ is a diagonal matrix
$\Delta_{\sigma}={\rm diag}(0,\cdots,0,{\rm e}^{2a\sigma(u'_i-u_i)}-1,0,\cdots)$.

\twocolumn
Since the Monte Carlo samplings are generated with the weight $|w|$, the
expectation value is calculated from the summation with the sign of $w$
(denoted as sign$w$).
For example, the expectation value of nearest neighbor transfer term is given by
\begin{equation}
<c^{\dag}_{i\sigma}c_{j\sigma}>= \sum_m (G_{\sigma})_{ji}{\rm sign}w
/\sum_m {\rm sign}w,
\end{equation}
where
\onecolumn
\begin{eqnarray}
(G_{\sigma})_{ji}&=& ({\rm exp}(-\lambda'' K^{\sigma})\cdots 
{\rm exp}(-\lambda K^{\sigma}){\rm exp}(V^{\sigma}(s,\alpha))\phi^{\sigma}_0
g \phi^{\sigma\dag}_0{\rm exp}(V^{\sigma}(u,\alpha))\nonumber\\
&\times&{\rm exp}(-\lambda K^{\sigma})\cdots{\rm exp}(-\lambda'' K^{\sigma}))_{ji}.
\end{eqnarray}
\twocolumn
$\sum_m$ denotes a summation over the Monte Carlo samples.
The two body correlation functions are similarly calculated using the Wick's
theorem.

In standard projector QMC approaches, we encounter an inevitable sign problem.  In our calculations it has
turned out that the negative sign problem is not serious, which enables us to 
estimate the
expectation values even in two space dimensions.

\section{Comparison with exact results and other QMC methods}

This section is devoted to examine the validity of our method by comparing 
energies obtained using the off-diagonal functions 
with the exact results for the 2D Hubbard model.
In Table I we compare our results
with the exact values and available data from quantum Monte Carlo method for 
the closed shell case.

\begin{table*}
\caption{Ground state energies for the 2D Hubbard model.
The boundary conditions are periodic in both directions. 
$\psi_0$ is set to be the normal free electron state.
The Constrained Path Monte Carlo (CPMC) results are from 
Ref.\protect\CITE{zha97}.
The projector Quantum Monte Carlo (QMC) results are from 
Ref.\protect\CITE{fur92}.
}
\centering
\begin{tabular}{cccccccccc}
\cline{1-10}
\cline{1-10}
  & $N_e$ & $U$ & $\psi_G$ & $\psi^{(2)}$ & $\psi^{(3)}$ & Extrapolated & CPMC & QMC & Exact\\
\cline{1-10}
4$\times$4 & 10 & 4 & -19.39 & -19.57 &  &  & -19.582 & -19.58 & -19.581 \\
4$\times$4 & 10 & 8 & -17.06 & -17.47 & -17.49  & -17.51  & -17.48  &        & -17.510 \\ 
10$\times$10 & 82 & 4 & -106.7 & -109.1 & -109.3 & -109.6 & -109.55 & -109.7 &\\
\cline{1-10}
\end{tabular}
\end{table*}

Obviously $\psi^{(m)}$ $(m=2,3)$ are improved appreciably compared with the
Gutzwiller function showing energies which are very close to the exact values.
Extrapolated energies are obtained through an extrapolation 
from the data obtained for
$\psi^{(0)}=\psi_{{\rm G}}$, $\psi^{(1)}$ , $\psi^{(2)}$ and $\psi^{(3)}$, as is shown
in Figs.1 and 2.
In the above calculations for the closed shell case, the lower level wave
function $\psi^{(2)}$ is found to be a nice approximation for the ground state
function.
In Table II we show results for the correlation functions comparing them
with the exact values.

\begin{figure}
\centerline{\psfig{figure=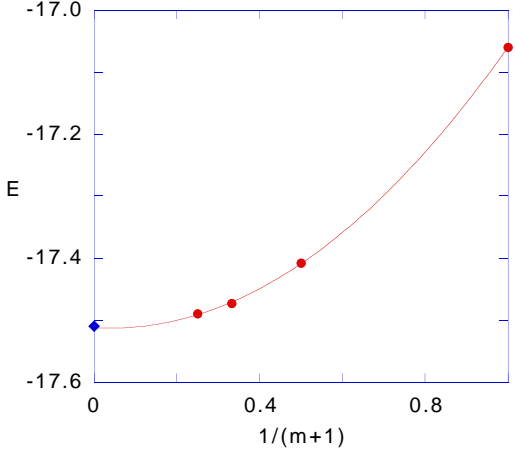,height=6cm}}
\caption{The energy for $N_e=10$ and $U=8$ on the $4\times 4$ lattice.
The results for $\psi_{{\rm G}}$, $\psi^{(1)}$, $\psi^{(2)}$ and
$\psi^{(3)}$ are shown. 
The diamond denotes the exact value.
}
\label{fig1}
\end{figure}

\begin{figure}
\centerline{\psfig{figure=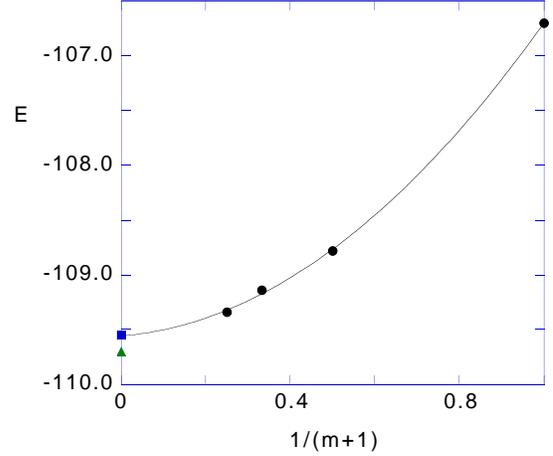,height=6cm}}
\caption{The energy for $N_e=82$ and $U=4$ on the $10\times 10$ lattice.
The results for $\psi_{{\rm G}}$, $\psi^{(1)}$, $\psi^{(2)}$ and
$\psi^{(3)}$ are shown.
The square denotes the
expectation value by CPMC and the triangle denotes that by QMC.
}
\label{fig2}
\end{figure}

\begin{table}
\caption{Correlation functions for the $4\times 4$ Hubbard model with
periodic boundary conditions in both directions. 
$N_e=10$, $U=4$ and $\psi_0$ is the normal free electron state.
}
\centering
\begin{tabular}{ccccc}
\cline{1-5}
      & $\psi_G$ & $\psi^{(2)}$ & CPMC & Exact \\
\cline{1-5}
$S(\pi,\pi)$     & 0.808 & 0.73  & 0.729 & 0.733  \\
$C(\pi,\pi)$     & 0.460 & 0.52  & 0.508 & 0.506  \\ 
$\Delta_{yy}(1)$ & 0.080 & 0.076 &       & 0.077  \\
$\Delta_{yy}(2)$ & 0.007 & 0.006 &       & 0.006  \\
$\Delta_{xy}(0)$ & 0.133 & 0.12  &       & 0.122  \\
$\Delta_{xy}(1)$ &-0.013 &-0.015 &       &-0.014  \\
$s(0,0)$         & 0.525 & 0.53  &       & 0.533 \\
$s(1,0)$         &-0.098 &-0.091 &       &-0.0911 \\
$c(0,0)$         & 0.33  & 0.33  &       & 0.326 \\
$c(1,0)$         &-0.047 &-0.056 &       &-0.0539 \\
\cline{1-5}
\end{tabular}
\end{table}

In Table II the correlation functions are defined by
\begin{equation}
s(i_x,i_y)= \langle(n_{0\uparrow}-n_{0\downarrow})(n_{i\uparrow}-n_{i\downarrow})\rangle~~{\rm for}~{\bf r}_i=(i_x,i_y),
\end{equation}
\begin{equation}
c(i_x,i_y)= \langle(n_0-\langle n_0\rangle)(n_i-\langle n_i\rangle)\rangle~~{\rm for}~{\bf r}_i=(i_x,i_y),
\end{equation}
where $n_i=n_{i\uparrow}+n_{i\downarrow}$.
$S({\bf q})$ and $C({\bf q})$ are Fourier transforms of $s(i_x,i_y)$ and
$c(i_x,i_y)$, respectively:  
\begin{equation}
S({\bf q})= \sum_i{\rm exp}(i{\bf q}\cdot {\bf r}_i)s({\bf r}_i),
\end{equation}
\begin{equation}
C({\bf q})= \sum_i{\rm exp}(i{\bf q}\cdot {\bf r}_i)c({\bf r}_i).
\end{equation}

\noindent
The superconducting correlation function
$\Delta_{\alpha\beta}(\ell)$ is defined by
\begin{equation}
\Delta_{\alpha\beta}(\ell)= \langle\Delta^{\dag}_{\alpha}(i+\ell\hat{x})
\Delta_{\beta}(i)\rangle,
\end{equation}
where $\Delta_{\alpha}(i)=
c_{i\downarrow}c_{i+\alpha\uparrow}-c_{i\uparrow}c_{i+\alpha\downarrow}$
($\alpha=\hat{x}$ or $\hat{y}$; $\hat{x}$ and $\hat{y}$ indicate an unit
vector in x and y direction, respectively).  The expectation values of
correlation functions agree with the exact values considerably well.
In these calculations $\psi^{(2)}$ satisfactorily reproduces the correlation 
functions
as well as the ground state energy.
In Table II,
parameters are given by $g=0.30$, $\lambda=0.07$, $g'=0.36$ and
$\lambda'=0.13$ (for $U=4$).

Now let us show an another example where we consider the $4\times 4$ lattice
with 14 electrons for the periodic boundary condition in both directions.  
A famous sign problem occurs in this case for which the standard projector
QMC method gives poor estimates of energy for large $U$.
The quantum number of total momentum for $\psi^{\sigma}_0$ can take
$(0,\pi)$ and $(\pi,0)$ as well as (0,0).  Two kinds of initial trial states
$\psi_0$ are examined; one is the Fermi sea with zero total momentum for both
spin states and the other one has non-zero total momentum
$(0,\pi)$ for each spin state.
We incorporate the antiferromagnetic order into a trial wave function in the
latter case since the energy gain is appreciable.
In Table III we show our data.  The extrapolated values are obtained as shown
in Figs.3 and 4 where the solid curve and dotted curve indicate the
energies obtained for SDW and normal initial states, respectively.
Our calculations are supported by the property that the extrapolated values
for different initial states coincides with each other within statistical
errors.  The energy obtained for $\psi^{(3)}$ is comparable with that by
CPMC for large $U$ and extrapolated values are fairly close to exact values.

\begin{figure}
\centerline{\psfig{figure=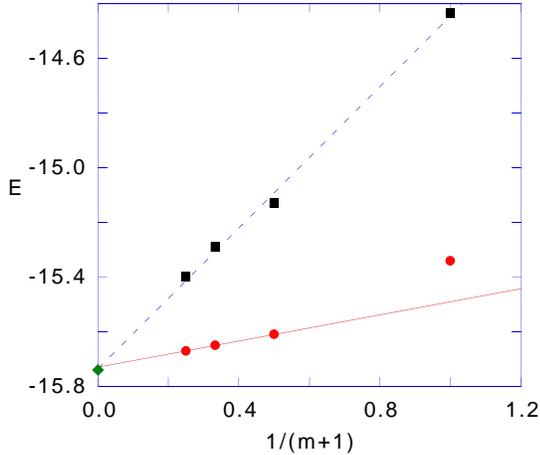,height=6cm}}
\caption{The energy as a function of $1/(m+1)$ for $N_e=14$ and $U=4$ on
$4\times 4$.
For the upper and lower curves an initial
function $\psi_0$ is set to be the Fermi sea or SDW state, respectively.
The diamond indicates the exact value. 
}
\label{fig3}
\end{figure}

\begin{figure}
\centerline{\psfig{figure=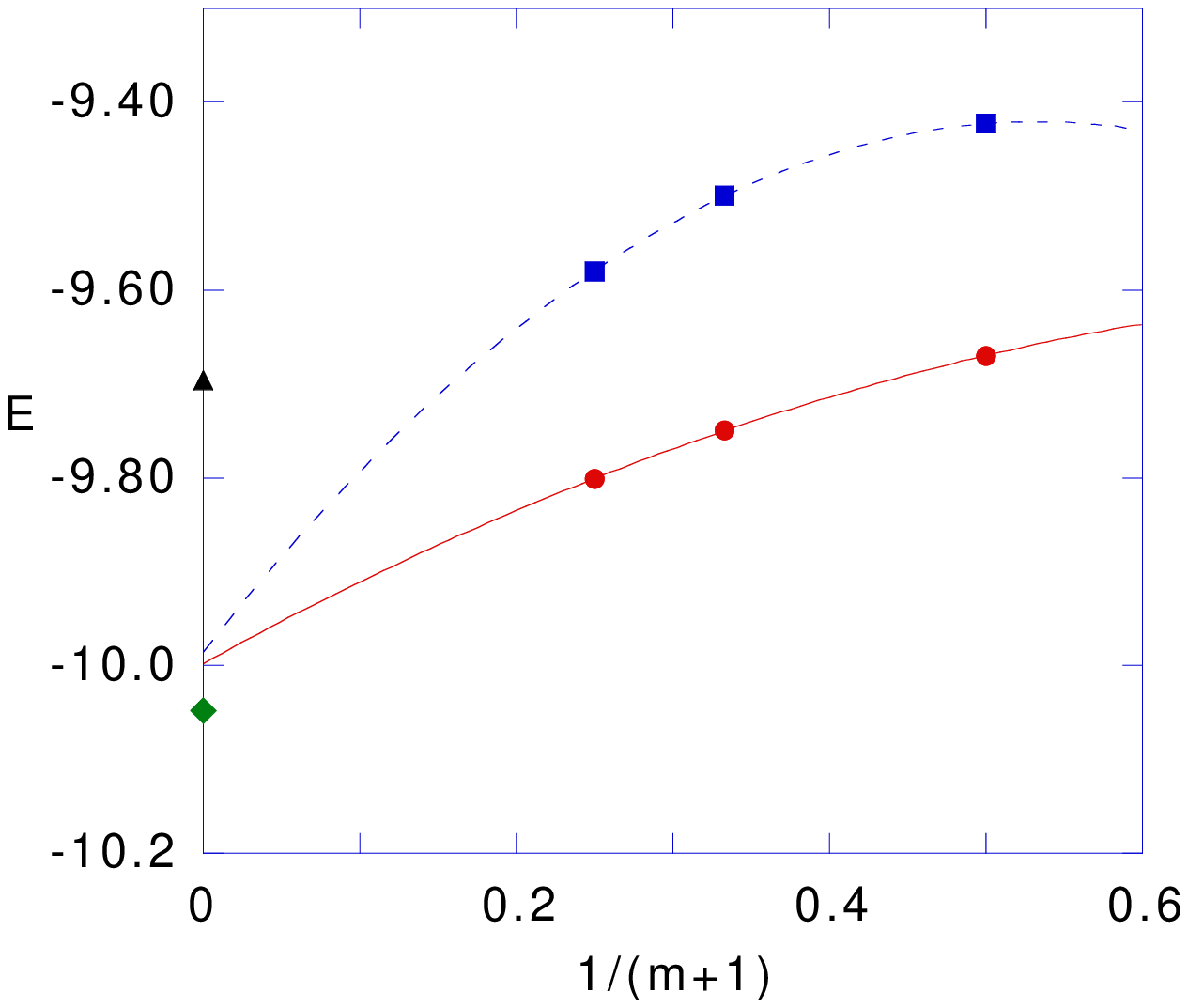,height=5.5cm}}
\caption{The energy as a function of $1/(m+1)$ for $N_e=14$ and $U=12$ on
$4\times 4$.
For the upper and lower curves an initial
function $\psi_0$ is set to be the Fermi sea or SDW state, respectively.
The diamond indicates the exact value and the triangle indicates CPMC result.
}
\label{fig4}
\end{figure}

\begin{table*}
\caption{Ground state energies for the $4\times 4$ Hubbard  with $N_e=14$
for the periodic boundary conditions in both directions, where $U=4$ and $U=12$.
$\psi_0$ is the non-interacting Fermi sea (normal) or SDW state (SDW) as 
indicated.
The exact diagonalization results are from Ref.\protect\CITE{par89}.
}
\centering
\begin{tabular}{cccccccc}
\cline{1-8}
  $U$ & $\psi_0$ & $\psi^{(0)}$ & $\psi^{(2)}$ & $\psi^{(3)}$ & Extrapolated & C
PMC & Exact \\
\cline{1-8}
4  & normal & -14.435  & -15.29 & -15.40 & -15.73 & -15.73 & -15.74 \\
4  & SDW    & -15.34   & -15.65 & -15.67 & -15.73 &        & -15.74 \\
12 & normal & -7.80    & -9.50  & -9.58  & -9.98  & -9.696 & -10.048\\
12 & SDW    &          & -9.75  & -9.80  & -10.0  &        & -10.048\\
\cline{1-8}
\end{tabular}
\end{table*}

\section{Results for the half-filled case}

In this section the 2D square lattice with the half-filled  band is investigated
as a first step toward more general cases.
The half-filled band systems have been studied numerically by the QMC
simulation\cite{hir85b} and VMC calculations.\cite{oht92,yok87b}
In our calculations
we impose the periodic boundary condition in one direction and antiperiodic
condition in the other direction to have a unique ground state.
Two sizes, $6\times 6$ and $8\times 8$, are investigated in this section.

The energy expectation values estimated by the off-diagonal wave function
Monte Carlo method (OWMC) are shown as a function of $1/(m+1)$
for $6\times 6$ in Fig.5.  The solid curves are drawn by the least squares
method for three data points obtained by $\psi^{(1)}$, $\psi^{(2)}$ and $\psi^{(3)}$.
The extrapolated values are used to obtain the total energy as a function
of $U$ as shown in Fig.6 for $6\times 6$.  In Fig.6 the energies for
$\psi_{{\rm G}}=P_{{\rm G}}\psi_0$, $P_{\rm G}\psi_{{\rm SDW}}$, $\psi^{(2)}$ 
and extrapolated values
are shown together.  For the Gutzwiller function the energy is greatly lowered
when the antiferromagnetic order is introduced.  We obtain lower energies
than those for $P_G\psi_{{\rm SDW}}$ by using the off-diagonal wave functions.
In order to find optimum parameters we use the correlated measurements
method\cite{umr88,kob93} to find the most descendent direction in the
parameter space.  We shift parameters along that direction starting from
a set of parameters.  We show our parameters in Table IV.
The extrapolations for $8\times 8$ as a function of $1/(m+1)$ are shown in
Fig.7 where the upper and lower curves correspond to $U=8$ and $U=4$,
respectively. 
The energy expectation values for $8\times 8$ are also shown in Fig.8 where the
energies  by $\psi_{{\rm G}}$, $P_G\psi_{{\rm SDW}}$, $\psi^{(2)}$ and
extrapolation are
compared to one another.  The variational energy of $\psi^{(3)}$ is very
close to that of $\psi^{(2)}$ and the extrapolated value.  The results by the
quantum Monte Carlo simulation by Hirsch are also shown as a reference.
Although the QMC gives slightly higher energy for $U=8$,
a good agreement between two calculations support our method as well as the
QMC simulation.  It is fortunate that the total energy for the intermediate
value of $U$ is almost independent of the system size, which means that
the extrapolated energies per site for $6\times 6$ and $8\times 8$ coincide 
with each other.
Thus the energies in Figs.6 and 8 are expected to agree
with the behavior of infinite systems.

\begin{figure}
\centerline{\psfig{figure=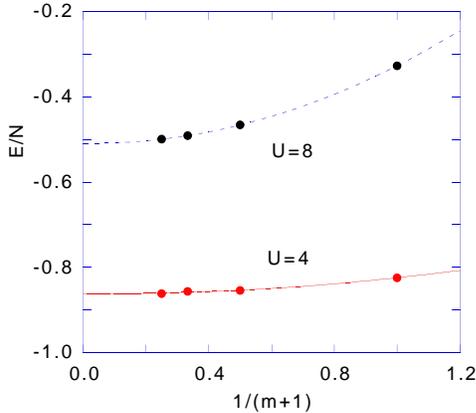,height=5.5cm}}
\caption{The energy as a function of $1/(m+1)$ for $6\times 6$ at
half-filling.
The upper and lower curves correspond to $U=8$ and $U=4$, respectively.
}
\label{fig5}
\end{figure}

\begin{figure}
\centerline{\psfig{figure=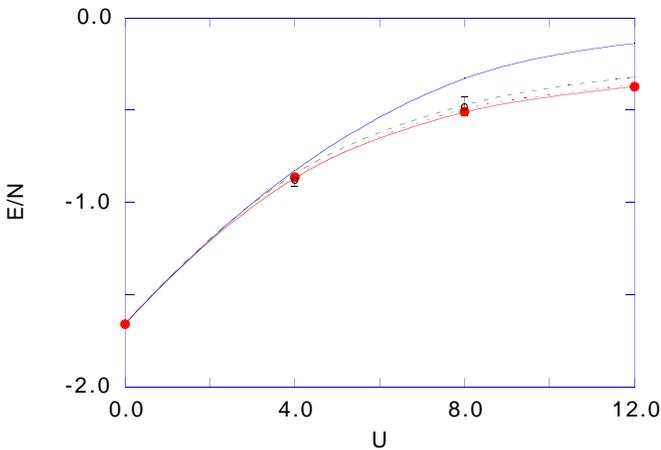,height=6cm}}
\caption{The energy as a function of $U$ for $6\times 6$ at half-filling.
From the top the energies for $\psi_{{\rm G}}$,
$P_{{\rm G}}\psi_{{\rm SDW}}$, $\psi^{(2)}$ and extrapolated values are shown.
The QMC results for $8\times 8$ are shown by open circles as a reference. 
}
\label{fig6}
\end{figure}

\begin{figure}
\centerline{\psfig{figure=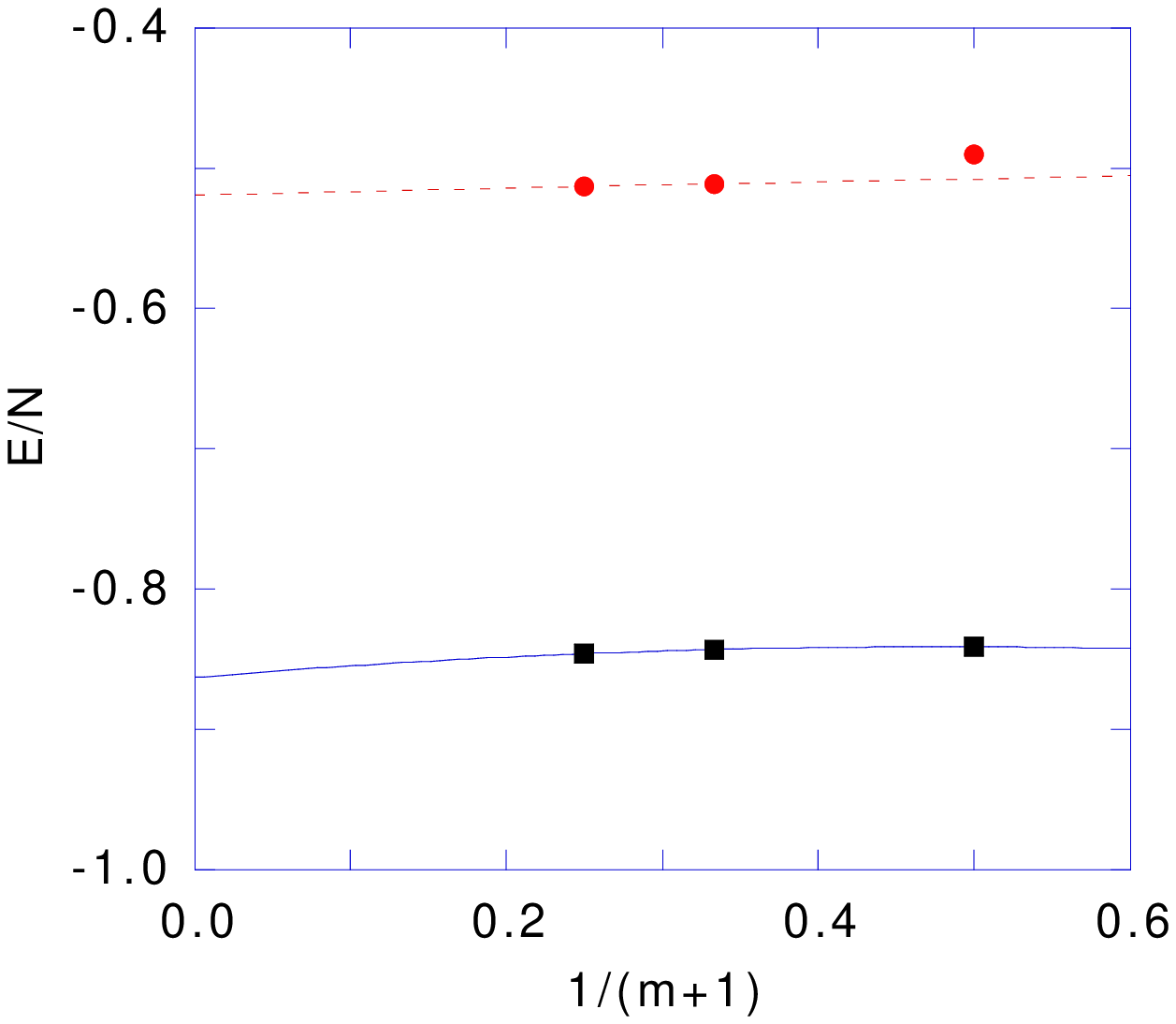,height=6cm}}
\caption{The energy as a function of $1/(m+1)$ for $8\times 8$ at half-filling. 
The upper and lower curves correspond to $U=8$ and $U=4$, respectively.
}
\label{fig7}
\end{figure}

\begin{figure}
\centerline{\psfig{figure=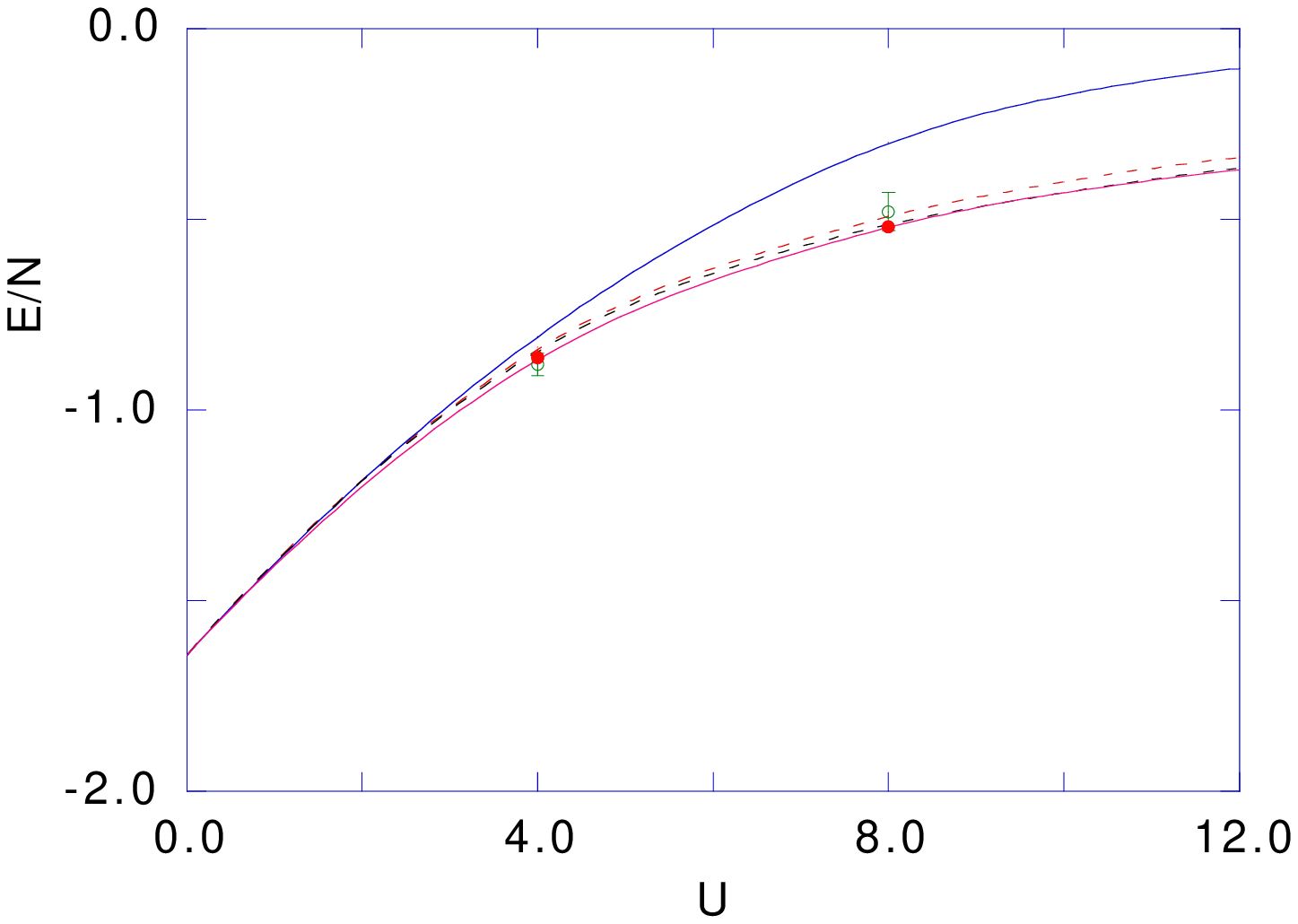,height=6cm}}
\caption{The energy as a function of $U$ for $8\times 8$ at half-filling. 
From the top the energies for $\psi_{{\rm G}}$,
$P_{{\rm G}}\psi_{{\rm SDW}}$, $\psi^{(2)}$ and extrapolated values are shown.
The QMC results are shown by open circles.
}
\label{fig8}
\end{figure}

\begin{table*}
\caption{Variational parameters used in the simulation for $6\times 6$.
}
\centering
\begin{tabular}{cccccccc}
\cline{1-8}
$U$ & wave function & $g$ & $\lambda$ & $g'$ & $\lambda'$ & $g''$ & $\lambda''$ \\
\cline{1-8}
4 & $\psi_{{\rm G}}$ & 0.55 & 0.0   & 1.0 & 0.0 & 1.0 & 0.0 \\
  & $\psi^{(1)}$     & 0.18 & 0.118 & 1.0 & 0.0 & 1.0 & 0.0 \\
  & $\psi^{(2)}$     & 0.47 & 0.00064 & 0.44 & 0.123 & 1.0 & 0.0 \\
  & $\psi^{(3)}$     & 0.40 & 0.105  & 0.38  & 0.126 & 0.398 & 0.128 \\
\cline{1-8}
8 & $\psi_{{\rm G}}$ & 0.24 & 0.0   & 1.0 & 0.0 & 1.0 & 0.0 \\
  & $\psi^{(1)}$     & 0.064 & 0.103 & 1.0 & 0.0 & 1.0 & 0.0 \\
  & $\psi^{(2)}$     & 0.17 & 0.00247 & 0.14 & 0.103 & 1.0 & 0.0 \\
  & $\psi^{(3)}$     & 0.18 & 0.159  & 0.128  & 0.197 & 0.142 & 0.096 \\
\cline{1-8}
\end{tabular}
\end{table*}

The spin structure factor defined as
\begin{equation}
S({\bf q})= \frac{1}{N} \sum_{j\ell}{\rm exp}(i{\bf q}\cdot ({\bf r}_j-{\bf r}_{\ell}))\langle(n_{j\uparrow}-n_{j\downarrow})(n_{\ell\uparrow}-n_{\ell\downarrow})\rangle,
\end{equation}
can also be evaluated with our method,
where $N$ denotes the total number of sites.  We show $S({\bf q})$ calculated
by $\psi^{(2)}$ and $\psi^{(3)}$ for $U=8$ in Figs.9 and 10, respectively,
as a function of $q_x$ and $q_y$.
The strong antiferromagnetic feature is shown in the figures with highly
enhanced component at ${\bf q}=(\pi,\pi)$.

\begin{figure}
\centerline{\psfig{figure=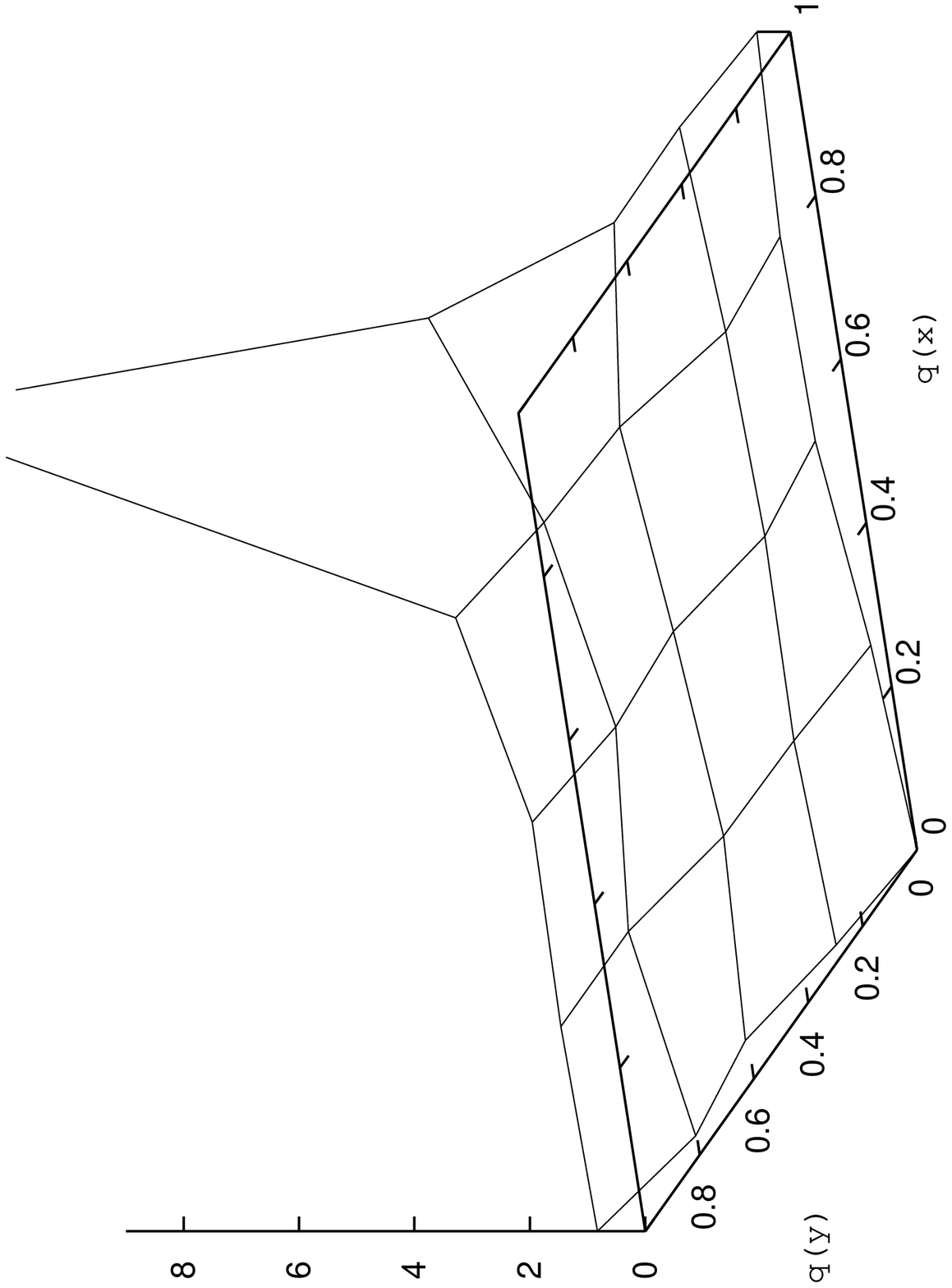,height=8cm}}
\caption{The spin structure factor for $\psi^{(2)}$.  $U=8$ for $8\times 8$
where $q(x)=q_x/\pi$ and $q(y)=q_y/\pi$.
}
\label{fig9}
\end{figure}

\begin{figure}
\centerline{\psfig{figure=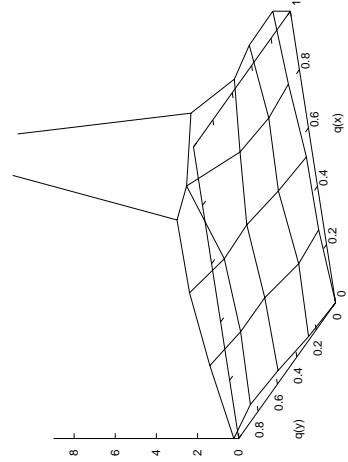,height=8cm}}
\caption{The spin structure factor for $\psi^{(3)}$.  $U=8$ for $8\times 8$
where $q(x)=q_x/\pi$ and $q(y)=q_y/\pi$.
}
\label{fig10}
\end{figure}

The sublattice magnetization $m$ has been calculated as a function of $U$
with the optimum variational parameters.  The sublattice magnetization is
defined as
\begin{equation}
m= \frac{1}{N}|\sum_j(-1)^j \langle n_{j\uparrow}-n_{j\downarrow}\rangle|
\end{equation}
$m$ is plotted in Fig.11 as a function of $U$ where the calculated values
obtained by $\psi^{(3)}$ (for $6\times 6$ and $8\times 8$), 
$P_{{\rm G}}\psi_{{\rm AF}}$\cite{yok87b} are shown comparing with 
QMC method\cite{hir85b}.
Our results are substantially lower than the VMC predictions with
the antiferromagnetically ordered state
$P_{{\rm G}}\psi_{{\rm AF}}$ and show a good agreement with QMC results.
The spin fluctuations appreciably reduce the sublattice magnetization but
do not destroy the order in two dimensions.

\section{Summary}
We have presented a new variational quantum Monte Carlo method for the
Hubbard model.  Our method is based on the property that we can systematically 
improve a variational wave function starting from the Gutzwiller wave
function. It is remarkable that the variational energy is
greatly lowered by the Gutzwiller and off-diagonal projection operators.
For the $4\times 4$ lattice our results agree remarkably well with 
the exact values obtained by the exact diagonalization.  Our calculations 
include the
case where a sign problem occurs for the standard projector QMC.  
The results for $4\times 4$ and $10\times 10$ (in Table I)
suggest that lower level off-diagonal functions show up
as good a level of approximations as the constrained path QMC. 
The extrapolated values are favorably compared with exact results.
For the 2D half-filled band system, we have calculated the exact energy
as a function of $U$, which is lower than that by the antiferromagnetically
ordered Gutzwiller function and mostly independent of the system size.
There is a good agreement with the results by the QMC simulation
concerning the energy and spin structure factor $S({\bf q})$. 
Thus we have demonstrated that the off-diagonal wave function Monte Carlo
(OWMC) is useful to investigate the ground state for strongly correlated
electrons.

In our simulation we spent most of time to search variational parameters
optimizing the energy expectation values.  
We use the correlated measurements method to find a most descendent direction
along which we shift the variational parameters starting from a set of
initial parameters.
Once the optimized parameters are
determined, the expectation values are estimated with a large number of 
Monte Carlo steps.

In the process of OWMC we must evaluated 
$\langle c^{\dag}_{i\sigma}c_{j\sigma}\rangle$ to estimate the kinetic energy,
which is the second process costing a lot of time in the simulation.
We feel that there is a room to improve our algorithm on this point.

Following a first step toward a development of the variational theory based
on the off-diagonal wave functions performed in this paper, we are planning
to consider some problems for correlated electrons.  In particular,
a possibility of the superconductivity for the non-half-filled band is    
considered to be an interesting one.  An application to other models also 
deserves an investigation.

We thank Professor H. Aoki and Dr. K. Kuroki for useful comments and 
discussions.
We express our sincere thanks to Professor P. Fulde for his suggestion
about considering exp-$S$ type functions in the simulation.
Our computations were supported by Research Information Processing System
(RIPS) at the Agency of Industrial Science and Technology (AIST) in Tsukuba.

\begin{figure}
\centerline{\psfig{figure=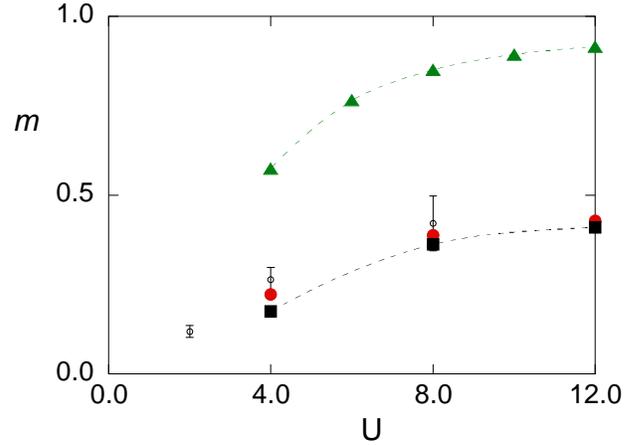,height=6cm}}
\caption{The sublattice magnetization as a function of $U$.  The results by
$\psi^{(3)}$ (solid circles for $6\times 6$ and squares for $8\times 8$), 
$P_{{\rm G}}\psi_{{\rm AF}}$ (triangles) and QMC 
(open circles)\protect\cite{hir85b}.
}
\label{fig11}
\end{figure}

\end{document}